\documentclass[aps,prl,twocolumn,superscriptaddress,10pt,longbibliography]{revtex4-2} 

\usepackage{amsmath}
\usepackage{amssymb}
\usepackage{commath}
\usepackage{mathtools}
\usepackage{braket}
\usepackage{dsfont}
\usepackage{bm}
\usepackage{orcidlink} 
\usepackage{graphicx}
\usepackage[capitalise]{cleveref}

\usepackage{hyperref}
\hypersetup{
  colorlinks=true,
  allcolors = blue,
  pdftitle={}
}

\newcommand{\scref}[2]{\namecref{#1}~\hyperref[#1]{\ref*{#1}#2}}

\newcommand{\sect}[1]{{\it #1---\!}}

\DeclareMathOperator*{\tr}{\rm tr}
\DeclareMathOperator*{\im}{\rm Im}
\DeclareMathOperator*{\hc}{h. c.}
\DeclarePairedDelimiter\ev{\langle}{\rangle} 

\begin{document}

\title{Kondo Echo Dynamics of Terahertz-Pumped Heavy Fermions}

\author{Francisco Meirinhos\,\orcidlink{0000-0002-3864-7569}}
\affiliation{\hbox{Physikalisches Institut and Bethe Center for Theoretical Physics, Universit\"at Bonn, Nussallee 12, 53115 Bonn, Germany}}

\author{Michael Turaev\,\orcidlink{0009-0003-8566-5078}}
\email{mturaev@uni-bonn.de}
\affiliation{\hbox{Physikalisches Institut and Bethe Center for Theoretical Physics, Universit\"at Bonn, Nussallee 12, 53115 Bonn, Germany}}

\author{Michael Kajan\,\orcidlink{https://orcid.org/0009-0002-5250-0405}}
\affiliation{\hbox{Physikalisches Institut and Bethe Center for Theoretical Physics, Universit\"at Bonn, Nussallee 12, 53115 Bonn, Germany}}

\author{Tim Bode\,\orcidlink{0000-0001-8280-3891}}
\affiliation{\hbox{Institute for Quantum Computing Analytics (PGI-12), Forschungszentrum J\"ulich, 52425 J\"ulich, Germany}}

\author{Johann Kroha\,\orcidlink{0000-0002-3340-9166}}
\email{jkroha@uni-bonn.de}
\affiliation{\hbox{Physikalisches Institut and Bethe Center for Theoretical Physics, Universit\"at Bonn, Nussallee 12, 53115 Bonn, Germany}}
\affiliation{\hbox{School of Physics and Astronomy, University of St.\,Andrews, North Haugh, St.\,Andrews, KY16 9SS, United Kingdom}}

\begin{abstract}
We provide a theoretical framework to describe the nonequilibrium temporal dynamics of correlated electron systems for realistic system parameters and the consequent, often exponentially long timescales. It is based on an entirely integrodifferential formulation of time-dependent dynamical mean-field theory, the noncrossing approximation, and the quantum representation of a driving electromagnetic field. For heavy-fermion systems, we identify two key nonequilibrium mechanisms governing their time evolution after a single-cycle terahertz excitation: transient, instantaneous shift from the Kondo toward the mixed-valence regime by an enhanced, photoassisted hybridization, and slow recovery of the heavy-fermion state due to the long Kondo coherence time. This explains recent time-resolved terahertz spectroscopy experiments microscopically and establishes the latter as a technique for direct experimental access to the Kondo coherence time and to the heavy-fermion quasiparticle weight, central for the classification of heavy-fermion quantum phase transitions. 
\end{abstract}

\maketitle

The past twenty-five years have witnessed a tremendous development of ultrafast spectroscopy techniques which have enabled both the creation and control of nonequilibrium quantum states of matter, as well as the measurement of system properties on ultrafast timescales \cite{Roadmap_2021}. While at first, weakly correlated phenomena were investigated, including ultrafast control of magnetism \cite{kirilyuk_ultrafast_2010,Matsubara_EuO_2015}, the melting of charge-density waves \cite{Wolf_CDWmelting_2008,Demsar_nonthermalCDW_2011}, and exciton dynamics \cite{Srivastava_2015,Staehler_2022}; more recently the focus has shifted to also include strongly correlated electron systems like cuprate superconductors \cite{fausti_light-induced_2011}, Mott-Hubbard systems \cite{Perfetti_Mott_2008,Wolf_2014}, and heavy-fermion (HF) compounds \cite{Lohneysen_HF_RMP2007,Demsar2006, Wetli_2018}.  The latter systems are peculiar, in that the Kondo electronic correlations generate a narrow, low-energy band at the Fermi energy, whose width is in the terahertz (THz) range \cite{Lohneysen_HF_RMP2007}. The investigation of their temporal dynamics has received a boost by the recent application of phase-sensitive THz time-domain spectroscopy \cite{Wetli_2018,Pal_2019,Yang_THzcond_2020,yang_critical_2023,Li_missing-weight_2025}, which has shown strikingly more complex behavior than expected: A HF system responds to an incident single-cycle THz pulse by emitting a time-delayed echo pulse, which can be interpreted to carry information on the strongly correlated dynamics, separated in time from the uncorrelated background \cite{Wetli_2018}.

On the theoretical side, real-time evolution methods have used a time-dependent density-matrix renormalization group \cite{gobert_real-time_2005,sentef_theory_2016,zhu_ultrafast_2021}, nonequilibrium real-time dynamical mean-field theory (t-DMFT) \cite{Georges_RMP_1996, Vollhardt_AnnPhys_2012,Eckstein_nonequilibrium_2010} with self-consistent diagrammatic \cite{Werner_2012,aoki_nonequilibrium_2014} or variational Monte Carlo \cite{ido_time-dependent_2015,fauseweh_laser_2020} impurity solvers, or a numerical renormalization group for quantum impurity problems \cite{anders_real-time_2005,weichselbaum_tensor_2012,nghiem_time-dependent_2014}. However, due to the complexity of correlated many-body systems, real-time evolution has largely been limited to quench dynamics \cite{freericks_quenching_2008,Werner_2012} and short evolution times \cite{fauseweh_laser_2020}. 
For the case of the echo-pulse response of HF systems \cite{Wetli_2018,Pal_2019}, a phenomenological rate-equation model could successfully explain the echo-pulse shape \cite{Wetli_2018}, but a microscopic understanding of important dynamical properties---most prominently the delay time---has remained elusive. 

The quantitative, microscopic time evolution of such a light-driven HF system requires (i) evolving the system from ultrashort timescales corresponding to the inverse binding energy of $4f$ electrons of $O(\mathrm{eV})$, to multiples of the inverse lattice Kondo temperature $T^*_{\rm K}$ of $O(\textrm{meV})$; (ii) treating the photon field as a dynamical quantum field, since it not only drives the correlated electron system, but has its own dynamics controlled by the coupling to the correlated electrons; and (iii) maintaining a thermalizing environment which ultimately restores the system to thermal equilibrium at the cryostat temperature $T$. Note that the electromagnetic environment cannot serve for this purpose because of insufficient photon spectral density. 
Here, we develop a formulation of the t-DMFT for a driven-dissipative Anderson lattice which incorporates all of these requirements. In particular, it describes both the strongly retarded evolution of the correlated lattice electrons and the absorption and emission dynamics of a THz photon field, far beyond the exponentially long Kondo coherence timescale $\tau_{\rm K}^*=2\pi\hbar/(k_{\rm B}T^*_{\rm K})$, where $\hbar$ and $k_{\rm B}$ are the Planck and the Boltzmann constants, respectively. This reveals the underlying microscopic mechanisms responsible for the collapse and delayed revival of the HF state observed in experiments \cite{Wetli_2018,Pal_2019}. The initial THz excitation provides an additional, photon-assisted hybridization channel between the rare-earth $4f$ orbitals and the conduction band. Our calculations show that this drives the system transiently from the Kondo to the mixed-valence regime, and thus causes an instantaneous collapse of the HF band even though the equilibrium Kondo coherence time $\tau_{\rm K}^*$ is a much longer timescale. 
Subsequently, the system relaxes to thermal equilibrium in a highly nonlinear fashion which involves the reconstruction of the correlated HF band, accompanied by the emission of a delayed, temporally confined burst of THz radiation after $\tau_{\rm K}^*$. Our analysis shows that this burst originates from the correlated-electron dynamics.

\sect{Model and methods} 
The Anderson lattice model for a generic HF compound with an additional dipole coupling to a photon mode reads
\begin{align}
\label{eq:hamiltonian}
    H & = -t_c \sum_{\langle i, j\rangle, \sigma} c_{i\sigma}^{\dagger}c_{j\sigma}^{\vphantom{\dagger}}  + \sum_{i, \sigma} \varepsilon_d^{\vphantom{\dagger}} d_{i\sigma}^{\dagger} d_{i\sigma}^{\vphantom{\dagger}} + {U}\sum_{i} d_{i\uparrow}^{\dagger} d_{i\uparrow}^{\vphantom{\dagger}} d_{i\downarrow}^{\dagger} d_{i\downarrow}^{\vphantom{\dagger}} \nonumber \\ 
    &+\Omega a^{\dagger}a +\left[ V - i g ( a - a^{\dagger} ) \right] \sum_{i,\sigma} \left( c^{\dagger}_{i\sigma}d^{\vphantom{\dagger}}_{i\sigma} + \hc \right).
\end{align}
Here $c_{i\sigma}^{\dagger}$, $c_{i\sigma}^{\vphantom{\dagger}}$ and $d_{i\sigma}^{\dagger}$, $d_{i\sigma}^{\vphantom{\dagger}}$ are the creation and destruction operators for electrons with spin $\sigma=\,\uparrow,\downarrow$ on lattice sites $i$ in the conduction band (typically $5d$) and in the rare-earth $4f$ orbitals, respectively. The conduction band is modeled by a semicircular density of states of half-bandwidth $D$, representative of a lattice with a nearest-neighbor hopping amplitude $t_c$. The $4f$ electrons have a binding energy $\varepsilon_d<0$ and a strong on-site repulsive interaction $U\gg |\varepsilon_d|$. The last term in \cref{eq:hamiltonian} describes the mixing between local $4f$ and conduction states by the intrinsic hybridization $V$ and due to the weak optical dipole coupling $g$ to a bosonic photon field $a^{\dagger}$, $a$ \cite{Dmytruk_2021,Li_2020}. Note that intraband transitions are suppressed by the dipole selection rules and that the photon field couples equally at all lattice sites. This is because the $f$-$c$ hybridization matrix elements are confined to a unit cell, and the photonic wavelength exceeds the lattice spacing by several orders of magnitude. The hybridizations $V$ and $g$ can be taken to be local in space, because a possible dependence on momentum transfer is integrated over in the effective $f$-$c$ coupling, so that this does not impair the accuracy of the model. Importantly, the photon quantum field has its own dynamics which we describe by an effective single mode with frequency $\Omega$ in the THz regime, comparable to the Kondo scale. This mode will obtain a homogeneous spectral broadening due to its ultrashort pulse duration, in accordance with the experiments \cite{Wetli_2018,Pal_2019,Yang_THzcond_2020,yang_critical_2023}. 
Treating the THz electromagnetic field in the quantized photon-number representation enables efficient, resonant excitations of electrons out of the HF band and distinguishing the intrinsic, Kondo coherent dynamics from superradiance, as shown below.

In contrast to previous formulations \cite{Eckstein_nonequilibrium_2010}, we compute all two-time 
Green functions by solving the Kadanoff-Baym (KB) integrodifferential equations, using the recently developed two-time adaptive time-stepping algorithm \cite{Meirinhos2022,Meirinhos2024,Bode2024}. While formally equivalent to iterating the Dyson integral equations, this substantially accelerates the computations, as the self-consistent DMFT conditions are fulfilled at each time step. As a state-of-the-art nonequilibrium impurity solver for t-DMFT, we employ the auxiliary-particle technique in the noncrossing approximation (NCA) \cite{Barnes1976,Coleman1984,Hettler1998,Eckstein_nonequilibrium_2010}, as it provides computationally efficient long-time evolution up to multiples of $\tau_{\rm K}^*$. For strong on-site repulsion, $U\gg |\varepsilon_d|$, the NCA is known to quantitatively reproduce the energy scales of the Anderson model \cite{Bickers_1985,Bickers_1987} and to correctly describe the equilibrium spectra down to energies well below $T_{\rm K}^*$ \cite{Ehm_PES-Exp_2007}, developing a spurious, nondivergent cusp only far below it. In the time domain, the latter results in an inaccuracy in the subleading long-time behavior, which, however, is washed out by the thermal bath discussed below. Thus, the NCA will reliably describe the long-time evolution and retardation effects of order $\tau_{\rm K}^*$ and beyond. We write the entire formalism in terms of the greater and lesser two-time propagators $G^{\gtrless}(t_1,t_2)$ for the matter and the photon fields, as this is most suitable for the NCA. The time-evolving HF spectra in terms of the quasiparticle energy $\omega$ computed below are obtained in the standard way by transforming the two-time propagators to Wigner center-of-motion (CoM) and relative times $t=(t_1+t_2)/2$, $\tau=t_1-t_2$, and Fourier-transforming with respect to $\tau$. 
 
It is important to note that in the long-time limit, thermal equilibrium at cryostat temperature is restored by coupling to a thermal bath. This may be composed of bosonic (e.g. phonon) excitations \cite{Demsar2006} or of the vast number of electrons not excited by the initial photon pulse \cite{Werner_2012}. We find that the fermionic bath is the more efficient thermalization channel because it operates via mere particle exchange, while the phonon-scattering mechanism is limited by a kinematic restriction, the Rothwarf–Taylor bottleneck \cite{Rothwarf_1967,Demsar2006}. Therefore, in the present Letter, we consider a fermionic bath with the following system-bath coupling Hamiltonian:
\begin{equation}
\label{eq:bath}
H_{SB}=\Gamma \sum_{i\sigma n}  (c_{i\sigma}^{\dagger} \gamma_{n\sigma}^{\vphantom{\dagger}} + \hc)\, ,
\end{equation}
where $\gamma_{n\sigma}$ are the fermionic bath operators. Integrating out the bath degrees of freedom leads to an additional conduction-electron self-energy $\Delta_{\mathrm{bath}}^{\gtrless}(t_1,t_2)=|\Gamma|^2 G^{\gtrless}_{\mathrm{bath}}(t_1,t_2)$. Here, the fermionic bath propagators
are obtained by Fourier-transforming the equilibrium frequency-dependent expressions, $G_{\rm bath}^{\gtrless}(\omega)=\mp i \pi \rho(\omega) f(\mp \omega)$, to the time domain $\tau$. Here, $f(\omega)$ is the equilibrium Fermi-Dirac distribution at the cryostat temperature and $\rho(\omega)$ is a featureless, broadband bath spectral density. By varying $\Gamma$, we make sure that the dissipative thermalization dynamics is slow enough not to influence the intrinsic Kondo dynamics, i.e., the delay time and line shape of the delayed pulse.

\begin{figure}[!t]
  \centering
  \includegraphics[width = \linewidth]{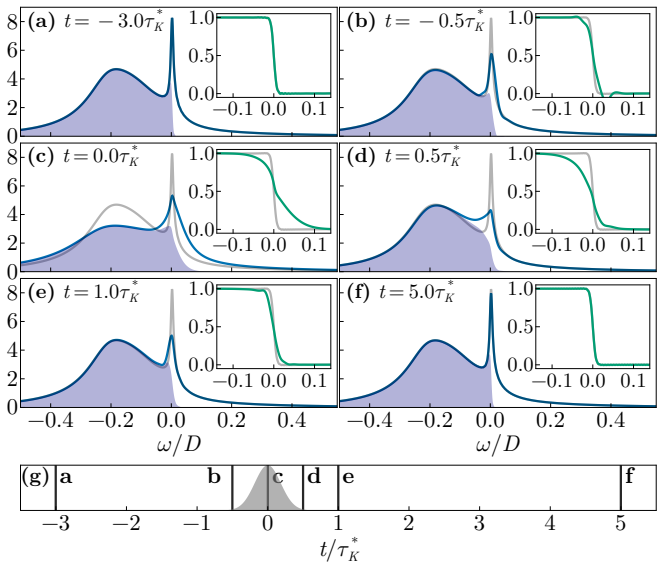}
  \caption{Time evolution of HF spectra due to an incident THz pulse. Panels (a)--(f) display the spectra at the CoM times marked by the corresponding letters in panel (g), which shows the timeline where the incident pulse envelope is displayed as a gray Gaussian peak, with $t = 0$ set at its maximum. Blue, solid lines correspond to momentum-integrated $4f$ spectral density $A_{d}(t,\omega)$, while shaded areas denote the occupied part of the spectra. The insets show the corresponding nonequilibrium distribution functions. The gray lines in the main panels and in the insets represent the equilibrium spectra and the Fermi-Dirac distribution, respectively, for comparison. The parameter values used are, in units of the conduction half-bandwidth D, $\varepsilon_d = -0.18\, D$, $V = 0.15\, D$, $U\to\infty$, and inverse temperature $\beta = 300/D$. The resulting lattice Kondo scale, extracted as the half-width of the equilibrium Kondo peak in (a), is $k_{\rm B}T^*_{\rm K}\approx 5 \times 10^{-3}\,D$. The THz driving pulse is characterized by $\Omega = 5 \times 10^{-4} \,D$, $\delta_0 = 67 \,D^{-1}$, $\overline{n}_a^{(0)} = 10$ (per $4f$ site), with $g = 0.02 \,D$ and bath coupling $\Gamma = 0.11 \,D$.}
  \label{fig:4f_spectra}
\end{figure}

\sect{Heavy-fermion dynamics}
To prepare the initial HF state in equilibrium, we solve for the steady state of the HF system [\cref{eq:hamiltonian}] in the presence of the thermal bath [\cref{eq:bath}] and thermal photon occupation; however, without an incident THz pulse. We truncate these functions at $t\ll 0$, retaining a sufficiently long history to preserve Kondo coherence while allowing for the subsequent incorporation of nonequilibrium dynamics induced by the THz pulse.
Memory truncation~\cite{schuler2018truncating} during time evolution is \emph{not} employed, as it would lead to spurious washing of correlated low-energy excitations. The incident two-point photon propagator, as produced in experiments by the nonlinear THz generation crystal with a Gaussian envelope of width $\delta_0$ \cite{Wetli_2018,Pal_2019}, acts as a driving field on the HF system and reads
\begin{equation}
    G^{(0) <}_{a}(t_1,t_2) = -i\, \overline{n}_a^{(0)} \,
    {\rm e}^{-(t/\delta_0)^2}\, {\rm e}^{-(\tau/2\delta_0)^2-i\Omega \tau}\, , 
    \label{eq:pump}
\end{equation}
where $\overline{n}_a^{(0)}$ is the total photon number in the incident pulse, and we have set the zero of CoM time $t$ at the maximum of the Gaussian envelope. Note that the Gaussian factor as a function of $\tau$ accounts for the fact that the incident THz pulse has a nonzero spectral width $1/\delta_0$ in the frequency domain. Within the local t-DMFT, this can equivalently be seen as due to an integral over a continuous range of photonic modes. \cref{fig:4f_spectra} shows the resulting time evolution of the momentum-integrated, total $4f$ spectrum and the occupied $4f$ spectral density of a HF system as well as the nonequilibrium distribution function, representative for a range of model-parameter values in the Kondo regime \cite{Meirinhos2024}. See \cite{Supplement} for a video of the continuous evolution. It exhibits two important phenomena. Long before the arrival of the THz pulse [\scref{fig:4f_spectra}{(a)}], the equilibrium spectrum shows the familiar, narrow Kondo resonance at the Fermi level and the broad single-particle resonance near $\omega\approx \varepsilon_d$, where most of the occupied spectral weight resides. At the onset of the THz pulse [\scref{fig:4f_spectra}{(b)}], the Kondo resonance starts to decline, while the $4f$ occupation number (gray area) is still essentially unchanged. However, at the peak of the THz pulse [\scref{fig:4f_spectra}{(c)}], a large fraction of the $4f$ spectral weight is shifted above the Fermi level, and hence, a large fraction of the $4f$ occupation number is moved to the conduction band. 
\begin{figure}[!t]
  \centering
  \includegraphics[width = \linewidth]{figs/Fig2.pdf}
  \caption{Time evolution of the energy- and momentum-resolved conduction spectral density (steep lines) and $4f$ spectral density (flat bands) of the THz driven Anderson lattice, $A_{\rm tot}(t,\omega)= - \im \sbr{G_{c_{\bm{k}}}(t, \omega) + G_{d_{\bm{k}}}(t, \omega)} / \pi$, in Wigner coordinates at the CoM times $t$ as indicated. These times correspond to the labels a,c,d,f in \scref{fig:4f_spectra}{(g)}, respectively. \emph{Top row}: Overall view. \emph{Bottom row}: Zoomed-in view on the Fermi level. The dependence on momentum $\bm{k}$ is displayed via the parameter $\varepsilon=\varepsilon_{\bm{k}}-\mu$, with $\mu$ being the chemical potential, since the bare $c$-electron dispersion $\varepsilon_{\bm{k}}$ enters the DMFT only through this parameter. Parameter values are as in \cref{fig:4f_spectra}.}
  \label{fig:HF_dispersion}
\end{figure} 
This is because of the additional, photoassisted hybridization channel induced by the incident pulse in \cref{eq:hamiltonian}, with an effective amplitude $\approx g\sqrt{\overline{n}_a}$ . It transiently shifts the system into the mixed-valence regime, so that the exponentially long coherence time $\tau_{\rm K}^*$ of the Kondo regime no longer exists, and the Kondo resonance is destroyed on a much shorter timescale by stimulated HF excitation. At the end of the THz pulse [\scref{fig:4f_spectra}{(d)}] it has almost completely collapsed. This is a much faster destruction mechanism than heating (thermal decoherence). At the same time, the distribution function (insets of \cref{fig:4f_spectra}) departs strongly from the equilibrium Fermi-Dirac distribution, indicating the creation of hot charge carriers. For subsequent times, in the absence of the THz radiation, the system returns to the Kondo regime. It thus takes the Kondo coherence time $\tau_{\rm K}^*$ for the Kondo resonance to start to build up [\scref{fig:4f_spectra}{(e)}] and an even longer time to completely recover [\scref{fig:4f_spectra}{(f)}]. 

The fast collapse and delayed recovery of the Kondo state is also seen in the HF band structure as calculated from our t-DMFT and shown in \cref{fig:HF_dispersion}. In the equilibrium band structure [\scref{fig:HF_dispersion}{(a)} and \hyperref[fig:HF_dispersion]{\ref*{fig:HF_dispersion}(f)}], the avoided crossing with an indirect gap between the conduction (steep line) and the flat HF band, 
as well as the broad $4f$ spectral density near $\varepsilon_d$, are clearly visible. At the peak of the THz pulse (c), both the HF band and the low-lying $4f$ single-particle band are strongly blurred. At the end of the pulse, the avoided crossing is completely destroyed, and the nearly free conduction band crossing the Fermi level is restored [\scref{fig:HF_dispersion}{(d)}, lower panel]. Interestingly, the destruction occurs predominantly near the avoided crossing, indicating the resonant THz excitation across the gap. 

 \begin{figure}[!t]
    \centering    \includegraphics[width=\linewidth]{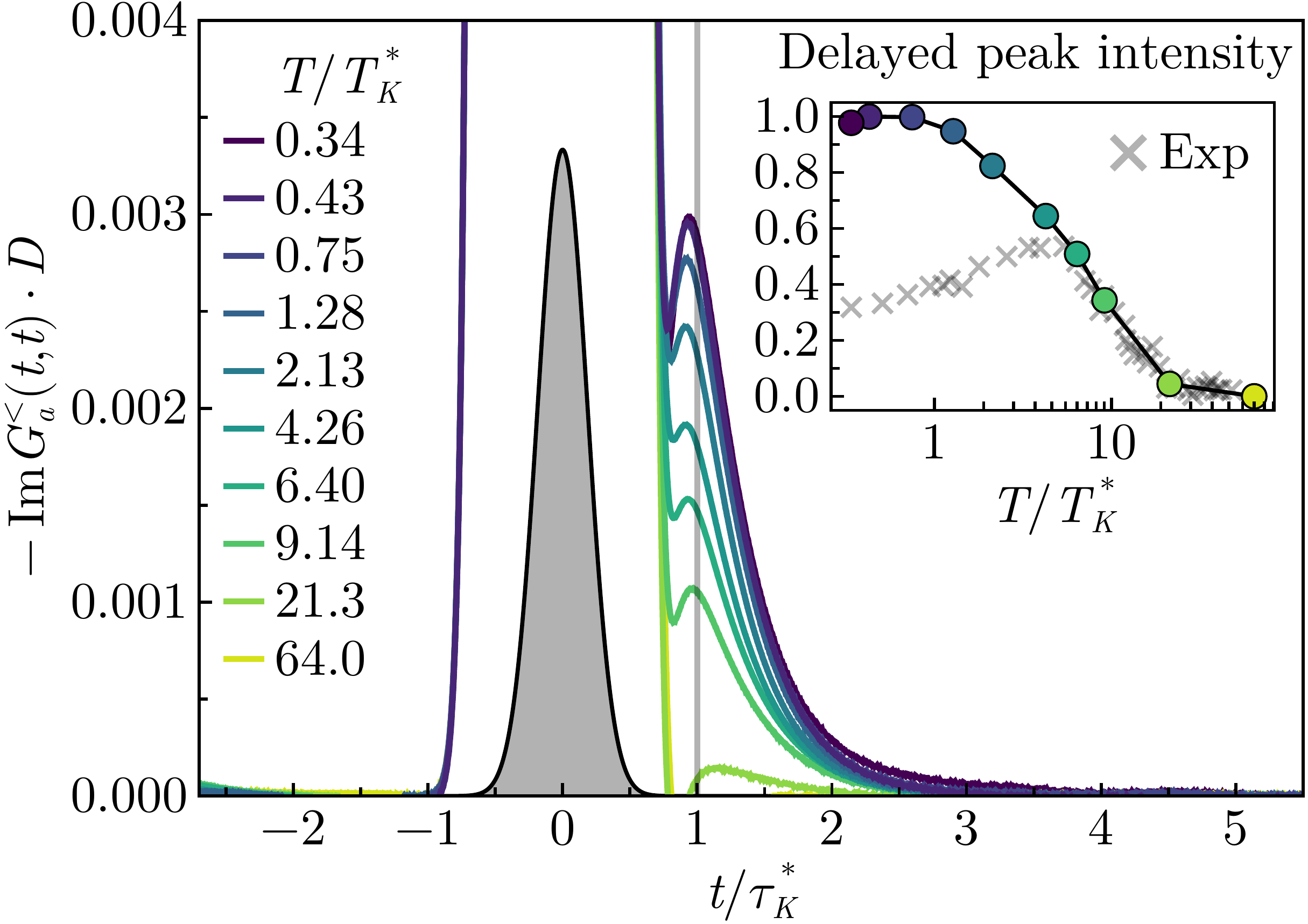}
    \caption{Time evolution of the renormalized photon intensity for various cryostat (bath) temperatures $T$, with the thermal photon occupation of the initial and the final states subtracted for clarity. The shaded area represents the incoming photon intensity downscaled by a factor of $3000$. The inset shows the $T$ dependence of the maximum intensity of the normalized delayed pulse (dots), compared to the normalized echo-pulse weight from the experiment in Ref.~\cite{Wetli_2018} (crosses). }
    \label{fig:Photon_emission_temp}
\end{figure}

\sect{Photon emission dynamics}
Photons are emitted by the delayed recombination of photoexcited electrons with holes in the recreated HF band. The photon number per $4f$ lattice site is given for the ingoing condition of the pump pulse [\cref{eq:pump}], as $n_a(t)=-\im G_a^<(t,t)$, where the renormalized, equal-time photon propagator $G_a^<(t,t)$ is computed by integrating its KB equation coupling to the matter Green functions $G_{c,d}^{\gtrless}$, cf. \cref{eq:DPhoton}.  The time evolution of $n_a(t)$ is shown in \cref{fig:Photon_emission_temp} for various temperature values $T$. It consists of two pulses: an intensive, instantaneous pulse followed by a substantially weaker one, delayed by the Kondo time $\tau_{\rm K}^*$. This is consistent with the experimental observations \cite{Wetli_2018,Pal_2019}, where the envelope of the instantaneous reflection is also significantly stronger even though it captures only a small space angle of the total emitted photon number. Downscaling the instantaneous pulse (incident photon number) to approximately the same height as the delayed one (gray area in \cref{fig:Photon_emission_temp}) shows that both pulses are well separated in time. To understand the computed time traces more quantitatively, we fit them by the sum of individual pulses. While the dominating instantaneous pulse has a Gaussian shape, we find a $1/\cosh^2[2\pi(t-\tau_{\rm K}^*)/w]$ pulse shape of delay time $\tau_{\rm K}^*$ and width $w$ for the delayed emission (see the End Matter), in agreement with the experimental findings and the phenomenological model of Ref.~\cite{Wetli_2018}. Note that this pulse shape corresponds to a Lorentzian-like line shape in frequency space. The change from an injected Gaussian to an emitted $1/\cosh^2$ pulse shape is a clear fingerprint of nonlinear dynamics. The inset of \cref{fig:Photon_emission_temp} shows the $T$ dependence of the maximum delayed-pulse intensity in comparison to the experimental findings for the HF compound CeCu$_{6}$~\cite{Wetli_2018}, scaled to the respective Kondo temperatures from theory and experiment. The logarithmic increase from high $T$ and settling at a constant value for $T\to 0$ is expected from the quasiparticle spectral weight in the heavy band and is in quantitative agreement with experiment down to about $5\,T^*_{\rm K}$. At lower $T$, the  experimental behavior deviates from this logarithmic increase. This is attributed to the proximity to the quantum phase transition in the doped compound CeCu$_{6-x}$Au$_x$ at $x=0.1$~\cite{Wetli_2018}, which is not included in our theory. The detailed agreement of our calculations in the range of applicability with the experiments strongly supports Kondo physics as the origin of the experimental findings.
\begin{figure}[!t]
    \centering
    \includegraphics[width = \linewidth]{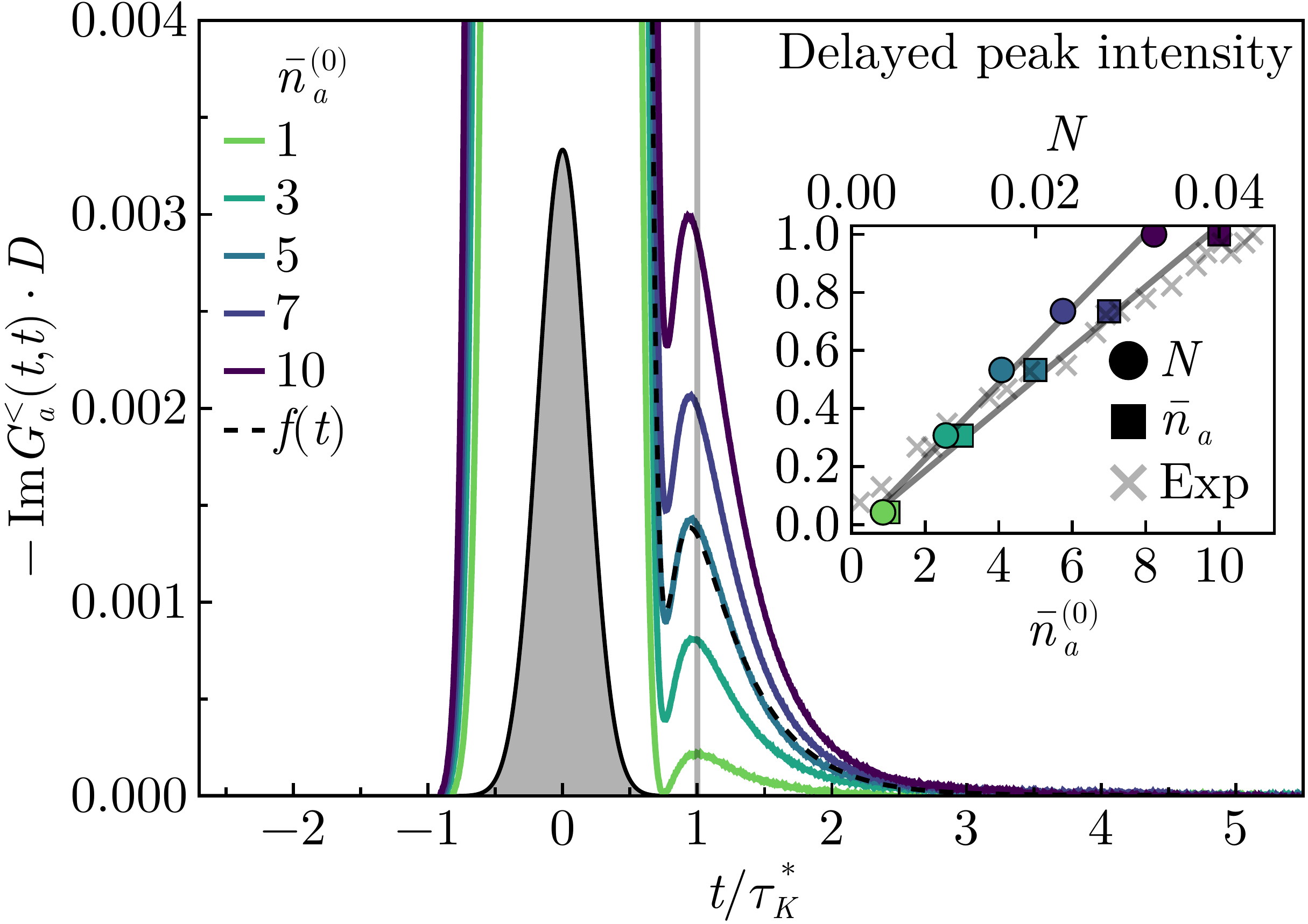}
    \caption{Renormalized photon intensity for varying incident pulse intensity $\overline{n}_a^{(0)}$. The thermal photon occupation in the initial and the final states has been subtracted. \emph{Inset}: The maximum intensity of the delayed pulse scales linearly with both  $\overline{n}_a^{(0)}$ and the fraction of excited electrons obtained by $N = \max_t \int_\omega [G^{>}_{c_{\sigma}}(t,\omega) +  G^{>}_{d_{\sigma}}(t,\omega)]$. A fit function $F(t) = F_0 / \cosh^2[2\pi (t-\delta t)/w]$, results in a delay time $\delta t=0.85\, \tau_{\rm K}^*$ and width $w=1.01\, \tau_{\rm K}^*$, in agreement with the Kondo interpretation \cite{Wetli_2018}.}
    \label{fig:Photon_emission_pump}
\end{figure} 

\sect{Kondo coherence versus superradiance}
It has been considered \cite{Yang_HF-super_2024} that the delayed photon pulse might be due to superradiance \cite{Dicke_SR_1954}, which, remarkably, would also have a $1/\cosh^2$ pulse shape \cite{Rehler_SR_1971}. In superradiance, quantum coherence among the emitting matter states is established by coupling to a common radiation mode, producing a photon burst with maximum intensity proportional to $N^2$, where $N$ is the number of excited matter states \cite{Rehler_SR_1971}. For HF systems, $N$ is proportional to the fraction of photoexcited electrons which, in turn, scales with the pump intensity $\overline{n}_a^{(0)}$ [cf. \cref{eq:pump}]. To conserve the injected and emitted energy, the pulse width and delay time would then scale as $1/\overline{n}_a^{(0)}$. Our theoretical formulation is capable of describing both the Kondo delay dynamics and superradiance, if it is realized, since the THz photon field is treated fully quantum-mechanically and couples identically to all lattice sites. To test superradiance, we plot the delayed pulse as a function of the pump intensity $\overline{n}_a^{(0)}\propto N$ in \cref{fig:Photon_emission_pump}. The figure clearly shows that the delayed peak intensity scales linearly with $N$ and the delay time is independent of $\overline{n}_a^{(0)}$. This indicates that superradiance is not realized in a generic HF system, presumably because the Pauli principle excludes multiple occupation of identical electron excited  states, and thus prevents quantum coherence between them, in contrast to distinguishable atoms.   

\sect{Conclusion} We have developed an efficient method that can describe the pumping of strongly interacting lattice electron systems by realistic, ultrashort optical or THz light pulses and evolve the systems well beyond the exponentially long timescales characteristic of strongly correlated electron systems. 
Treating the injected and emitted THz light as a dynamical quantum field allows us to directly compute the optical response observed experimentally \cite{Wetli_2018,Pal_2019}. For a generic heavy-fermion lattice system, the dynamics involve two processes: (i) instantaneous collapse of the Kondo state due to a transient shift to the mixed-valence regime by photoassisted $4f$ hybridization, and (ii) slow recovery of the heavy-fermion state on the scale of the lattice Kondo coherence time $\tau_{\rm K}^*$. To obtain these results, it is crucial to use the Anderson lattice model for the description, which includes the deeply bound $4f$ single-particle state and mixed-valence physics, in contrast to the Kondo model considered in previous studies \cite{Demsar2006,Werner_2012,fauseweh_laser_2020,zhu_ultrafast_2021}. At the same time, the pump frequency must be of the order of the Kondo scale, which corresponds to the THz range. Both processes, instantaneous collapse and delayed revival, provide the microscopic explanation for the experimental observations \cite{Wetli_2018}. This establishes the latter as a method to directly access the HF quasiparticle spectral weight as well as $\tau_{\rm K}^*$ \cite{Wetli_2018,Pal_2019,yang_critical_2023,Li_missing-weight_2025}. Our method is generally applicable to optically driven, correlated systems. We conjecture that similarly to heavy-fermion systems, charge fluctuations may also be important in other correlated systems even when the ultrafast THz field directly drives only the low-energy spin dynamics. Although superradiance is excluded for generic heavy-fermion compounds, the conditions for its realization in many-body systems can be explored in future work.
\vspace*{11pt}

\sect{Acknowledgements} We are grateful to Manfred Fiebig, Jingwen Li,  Hilbert v. L\"ohneysen, Alois Loidl, and Shovon Pal for insightful discussions. This work was funded by the Deutsche Forschungsgemeinschaft (DFG, German Research Foundation) under Germany's Excellence Strategy -- Cluster of Excellence Matter and Light for Quantum Computing, ML4Q (No. 390534769) and through the DFG Collaborative Research Center CRC 185 OSCAR (No. 277625399).

\vspace*{11pt}

\sect{Data availability} The code and data that support the findings of this article are openly available \cite{EchoPulseCode,Zenodo-data_2025}. 
 
\bibliography{refs}

\setcounter{equation}{0}
\setcounter{figure}{0}
\setcounter{table}{0}
\makeatletter
\renewcommand{\theequation}{A\arabic{equation}}
\renewcommand{\thefigure}{A\arabic{figure}}

\onecolumngrid
\newpage
\begin{center}
    \textbf{\large End Matter}
\end{center}
\vspace*{0.5cm}
\twocolumngrid

\sect{Integrodifferential formulation of t-DMFT} The t-DMFT formulation is expressed in terms of two-time Green functions for conduction electrons, $4f$ electrons, and photons, $G_{x}(t_1,t_2) = -i \,\ev{\hat{T}_{\mathcal{C}} \, \hat{x}(t_1) \hat{x}^{\dag}(t_2)}$, where $t_1,\,t_2$ lie on the Keldysh contour $\mathcal{C}$, and $\hat{T}_{\mathcal{C}}$ indicates contour time ordering.
For conduction electrons, $\hat{x}$ stands for $\hat{x}(t_1)=c(t_1)$; for $4f$ electrons, $\hat{x}(t_1)=d(t_1)$; and for photons, being a real field \cite{Flensberg2016}, $\hat{x}(t_1)=a(t_1)+a^{\dag}(t_1)$.

Within DMFT, the lattice problem is mapped onto an effective single-site problem embedded in an effective bath, described by an effectively noninteracting Green function, the Weiss field $\mathcal{G}$, which incorporates the remaining lattice under the self-consistency assumption that the Green functions on the impurity site are equal to the local (momentum-integrated) Green functions on each lattice site, and the impurity self-energy is local. For a conduction band with a semicircular density of states, the Weiss field satisfies \cite{aoki_nonequilibrium_2014}
\begin{align}
i \partial_{t_1} \mathcal{G}_{c}(t_1,t_2) = & \ \delta_{\mathcal{C}}(t_1,t_2) \label{eq:KBWeiss} \\ & + \sbr{\del {\frac{D^2}{4}G^{\rm loc}_{c} + \Delta_{\mathrm{bath}}} * \mathcal{G}_{c}}(t_1,t_2)~, \nonumber 
\end{align}
where $G^{\rm loc}_{c}$ is the local conduction-electron Green function, $\Delta_{\mathrm{bath}} = \Gamma^2 G_{\mathrm{bath}}$ denotes the hybridization with the fermionic bath, and $*$ denotes a contour-time convolution. Note that in our model, Eq.~(\ref{eq:hamiltonian}), only the conduction-electron component of the Weiss field is nontrivial, as they are the only ones with a hopping term. In the effective single-impurity Anderson model, the local conduction-electron Green function can be obtained via the T-matrix equation, $G^{\rm loc}_{c} = \mathcal{G}_{c} + \mathcal{G}_{c} * \mathcal{T}_c * \mathcal{G}_{c}$, where $\mathcal{T}_c$ is the conduction-electron T-matrix. By acting with $\mathcal{G}_{c}^{-1}$ we obtain an integrodifferential equation for the local conduction-electron Green function 
\begin{align}
    i \partial_{t_1} G^{\rm loc}_{c}&(t_1,t_2) = \delta_{\mathcal{C}}(t_1,t_2) \label{eq:KBGloc} \\ 
    + & \bigg( \mathcal{T}_{c} * \mathcal{G}_{c}  + \del{\frac{D^2}{4}G^{\rm loc}_{c} + \Delta_{\mathrm{bath}}} * G^{\rm loc}_{c}\bigg) (t_1,t_2).  \nonumber
\end{align}
For the Anderson model, we have $\mathcal{T}_{c}(t_1,t_2) =\Xi(t_1,t_2) \,G_{d}(t_1,t_2)$, where $\Xi(t_1,t_2) = V^2 + i g^2\,G_a(t_1,t_2)$ is the total $c$-$d$ hybridization function, dressed by interactions with the THz photons, and $G_{d}$ is the full, local $4f$-electron Green function which will be obtained using the NCA described in the next section. Note that \cref{eq:KBGloc,eq:KBWeiss} enforce the DMFT self-consistency condition at each time step \cite{Meirinhos2022} in that it uses the full lattice DMFT solutions at the previous times \cite{Meirinhos2024}. This KB integrodifferential formulation avoids additional DMFT iteration loops at each time step that are required in formulations based on Dyson equations \cite{Eckstein_nonequilibrium_2010,aoki_nonequilibrium_2014,zhu_ultrafast_2021}. 
The renormalized photon Green function is also computed via a T-matrix expression given by \cite{Meirinhos2024} 
\begin{align}
G_a(t_1,t_2) = G^{(0)}_{a}(t_1,t_2) + \del{G_{a}^{(0)} * \mathcal{T}_{a} * G^{(0)}_{a}}(t_1,t_2),   \label{eq:DPhoton}
\end{align}
where the photon T-matrix is
\begin{align}
    \begin{split}
        \mathcal{T}_{a}(t_1,t_2) = -i g^2 \sum_{\sigma} \big[ &G_{d\sigma}(t_1,t_2)\mathcal{G}_{c\sigma}(t_2,t_1) \\
        + &G_{d\sigma}(t_2,t_1)\mathcal{G}_{c\sigma}(t_1,t_2) \big].
    \end{split}
\end{align}

\sect{Nonequilibrium auxiliary-particle field theory}
In the limit of strong repulsion, $U\to\infty$, the constraint of no double occupancy of the local $4f$ orbital can be implemented faithfully by the auxiliary-particle representation \cite{Barnes1976,Coleman1984} of the $4f$ electron operators,  $d_{\sigma}^{\dagger}=f_{\sigma}^{\dagger}b$. Here, the fermionic operator $f_{\sigma}^{\dagger}$ creates a singly occupied $4f$ orbital with spin $\sigma$, the bosonic operator $b^{\dagger}$ creates the empty $4f$ orbital, and the operator constraint $\hat{Q}=\sum_{\sigma} f_{\sigma}^{\dagger}f_{\sigma}^{\vphantom{\dagger}} + b^{\dagger}b^{\vphantom{\dagger}} = \mathds{1}$ must be obeyed at all times. $\hat{Q}$ is a conserved quantity due to a local $U(1)$ symmetry in auxiliary-particle space. Since any physical operator, like $d_{\sigma}^{\dagger}$, annihilates the $Q=0$ sector of Hilbert space, the projector onto the physical sector $Q=1$ can be written as $\hat{P}_{Q=1}=\lim_{\lambda\to\infty}\mathrm{e}^{-\lambda(\hat{Q}-\mathds{1})}$, and the projected partition function reads in terms of the density matrix $\hat{\rho}$, $Z_{Q=1}=\langle \hat{\rho}\rangle_{Q=1} = \lim_{\lambda\to\infty} \tr\, [\hat{\rho} \mathrm{e}^{-\lambda(\hat{Q}-\mathds{1})} \hat{Q}] = \lim_{\zeta\to 0} \zeta^{-1} \langle \hat{Q} \rangle_{\zeta}$. Here, $\langle(\dots)\rangle_{\zeta}$ denotes the grand-canonical expectation value with respect to $\hat{Q}$, with the corresponding chemical potential $-\lambda$ and the fugacity $\zeta=\exp(-\lambda)$. This representation is exact for equilibrium \cite{Coleman1984} as well as for stationary \cite{Hettler1998} and time-dependent nonequilibrium situations \cite{Werner_2012,Bode2024}. As a consequence, the strongly correlated, constrained dynamics are achieved for any physical correlation function by first calculating it in the grand-canonical ensemble with respect to the auxiliary-particle number $\hat{Q}$ using standard field-theoretic techniques and then taking the limit $\zeta\to 0$. For any physical operator $\hat{A}$ acting on the impurity Hilbert space we thus have (see \cite{Bode2024} for details)
\begin{equation}
\ev{\hat{A}}
 = \lim_{\zeta\to0} \ev{\hat{A}}_\zeta / {\ev{\hat{Q}}_\zeta}\,.
 \end{equation} 
Note that the grand-canonical, greater ($>$) and lesser ($<$) auxiliary-particle propagators obey the $\zeta \to 0$ scaling, 
\begin{eqnarray}
G^{>}_{f\sigma,\, b}(t_1,t_2) &\sim& \mathcal{O}(1) \label{eq:zeta-scaling1}\\ 
G^{<}_{f\sigma,\, b}(t_1,t_2) &\sim& \mathcal{O}(\zeta)\, . \label{eq:zeta-scaling2}
\end{eqnarray} 
Because of this well-defined scaling behavior, it is convenient to formulate the entire t-DMFT formalism in terms of the greater and lesser Green functions. Additionally, the greater and lesser Green functions obey the symmetry $G^{\lessgtr}(t_1,t_2)^{*} = - G^{\lessgtr}(t_2,t_1)$. Therefore, once the Green function is time-evolved along the first time index, the evolution along the second index can be directly obtained from this relation. As a result, only a single diagonal step in the two-time plane remains, thus significantly simplifying the two-time evolution scheme.
\begin{figure}[t!]
    \includegraphics[width = \linewidth]{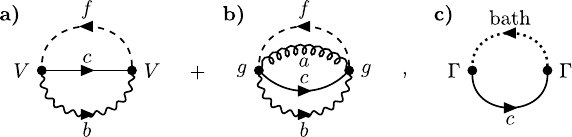}
    \caption{Diagrammatic representation of the NCA Luttinger-Ward functional,  composed of (a) the direct hybridization, (b) the photoassisted hybridization, and (c) the thermal bath. The solid lines represent the conduction electron, dashed lines the pseudofermion, wiggly lines the auxiliary boson, the curly line the THz photon, and the dotted line the bath propagators.}
\label{fig:NCA_diagrams}
\end{figure}

\sect{Noncrossing approximation} The Luttinger-Ward generating functional of the auxiliary-particle NCA is shown diagrammatically in \cref{fig:NCA_diagrams} and reads
 \begin{equation}
 \Gamma_2[G] = - \underset{\sigma}{\textstyle{\sum}}
   \underset{t_1, t_2}{\int} G_{c{\sigma}}(t_1,t_2) G_{b}(t_1,t_2) G_{f{\sigma}}(t_2,t_1) \Xi(t_1,t_2) 
\end{equation}
where all functions $G_{f,b,c,a}$ are understood as Keldysh matrix Green functions. The self-energies are obtained by functional derivatives, $\Sigma_{x}(t_1,t_2) = \pm i \, \delta \Gamma_{2}[G] / \delta G_{x}(t_2,t_1)$, $x=f\sigma,\, b,\, c\sigma,\, a$, where the upper and lower signs correspond to fermionic and bosonic self-energies, respectively. Since all the physical-particle self-energy diagrams ($c$ and $a$) are at least of $\mathcal{O}(\zeta)$, they vanish within NCA diagrams in the limit of $\zeta \to 0$. Thus, the physical propagators enter only in unrenormalized form into the equations of motion for the local problem. Using this, the equations of motion for the auxiliary-particle two-time greater and lesser Green functions read
 \begin{align}
    & (i\partial_{t_1} - E_{f,b}) G_{f,b}^{>}(t_1,t_2) = \int_{t_2}^{t_1}d\bar{t} \, \Sigma_{f,b}^{>}(t_1,\bar{t}) \, G_{f,b}^{>}(\bar{t},t_2) \label{eq:KBGNCAg} \\
    \begin{split}
    & (i\partial_{t_1} - E_{f,b}) G_{f,b}^{<}(t_1,t_2) = \int_{t_{0}}^{t_1}d\bar{t} \, \Sigma_{f,b}^{>}(t_1,\bar{t}) \,G_{f,b}^{<}(\bar{t},t_2) \\ & \phantom{(i\partial_{t_2} - \epsilon_{f,b}) G_{f,b}^{<}(t_1,t_2)} - \int_{t_{0}}^{t_2}d \bar{t} \, \Sigma_{f,b}^{<}(t_1,\bar{t})\, G_{f,b}^{>}(\bar{t},t_2), \label{eq:KBGNCAl} 
    \end{split}
\end{align}
where $E_{f} = \varepsilon_{f}$ for pseudofermions and $E_{b} = 0$ for auxiliary bosons. The auxiliary-particle self-energies \hbox{are given by}
\begin{align}
        & \Sigma_{b}(t_1,t_2) = -i\, \Xi(t_1,t_2)\sum_{\sigma} G_{f{\sigma}}(t_1,t_2)\,\mathcal{G}_{c{\sigma}}(t_2,t_1) \\
        & \Sigma_{f{\sigma}}(t_1,t_2) = i\, \Xi(t_1,t_2)\, \mathcal{G}_{c{\sigma}}(t_1,t_2)\,G_{b}^{\vphantom{0}}(t_1,t_2). 
\end{align}
Finally, the full, local $4f$-electron Green function is obtained within NCA as
\begin{align}
G_{d{\sigma}}(t_1,t_2) = \lim_{\zeta \to 0 }\frac{i\, G_{f{\sigma}}(t_1,t_2) G_{b}(t_2,t_1)}{\ev{Q}_{\zeta}}.
\end{align}
Thus, \cref{eq:KBWeiss,,eq:KBGloc,eq:KBGNCAg,eq:KBGNCAl,eq:DPhoton} constitute a complete set of integrodifferential equations, enabling long-time evolution with simultaneously high resolution of the system, \cref{eq:hamiltonian,eq:bath}. 

\begin{figure}[t!]
    \centering
    \includegraphics[width=0.75\linewidth]{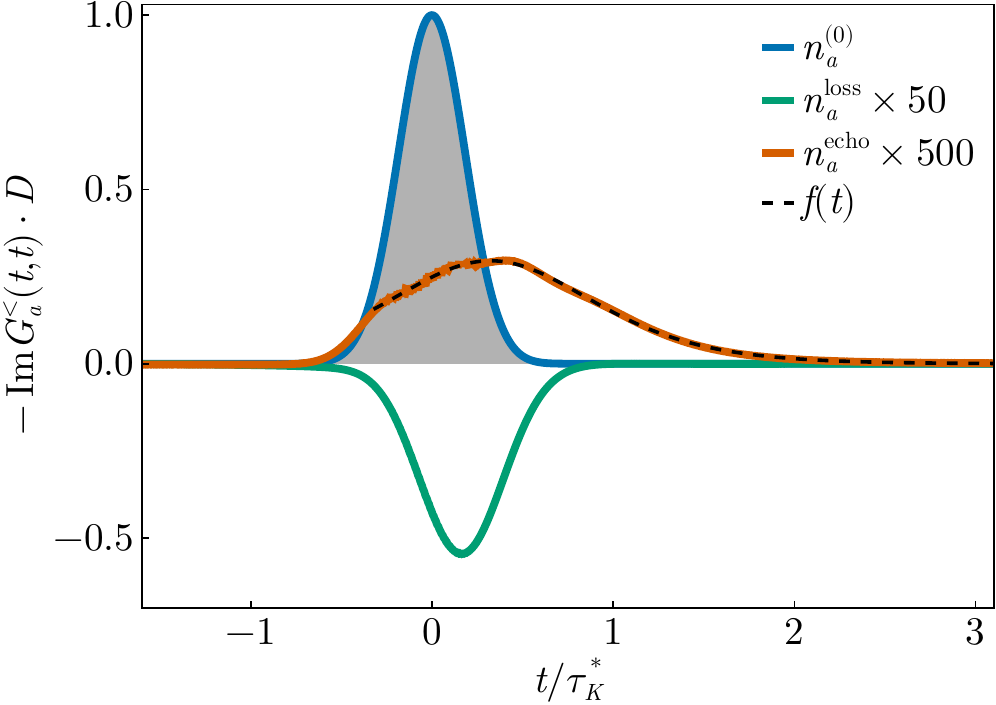} 
    \vspace*{-0.1cm}
    \caption{Fitting the time trace. For display, the curves shown are scaled by a factor as indicated. The green and the blue curves show the instantaneous response and the reduction of the response due to absorption by the bath, respectively, both following a Gaussian pulse shape. The red curve is what remains of the signal when these two contributions have been subtracted, displaying the relevant echo pulse. The best fit is obtained by a fit function $F(t)= F_0/\cosh^2[2\pi(t-\delta t)/w]$ to the echo pulse (dashed line), with fit parameters $F_0$, $\delta t$, and $w$. The sum of the red curve and the negative, blue curve produces the observable, delayed pulse seen in \cref{fig:Photon_emission_temp,fig:Photon_emission_pump}.}
    \label{fig:Fitting}
\end{figure} 
 
\sect{Fitting procedure} Due to the presence of additional dissipation channels coupled to the system, a direct fitting of only the initial pulse and the echo pulse is insufficient. The bath enables a radiationless relaxation to equilibrium, thus reducing the response of the renormalized photon pulse. To account for this, we allow for an extra Gaussian term with a negative amplitude in the fitting procedure. This three-component fit accurately reproduces the shape of the renormalized photon pulse, see \cref{fig:Fitting}. The width of the echo pulse extracted from the fit agrees with the Kondo timescale $\tau_{\rm K}^*$ to within 1\%, where $\tau_{\rm K}^*$ is determined as in Fig.~\ref{fig:4f_spectra}.  Notably, the height of the echo pulse is only about 0.1\% of the incoming pulse, indicating that only a small fraction of the photoexcited electrons undergo radiative relaxation, in agreement with the experimental observations \cite{Wetli_2018,Pal_2019}. The shape and the width of the echo pulse follows a $1/\cosh^2$ function in agreement with the phenomenological approach of Ref.~\cite{Wetli_2018}.

\vfill

\end{document}